\input harvmac

\noblackbox

\let\includefigures=\iftrue
\let\useblackboard=\iftrue
\newfam\black

\includefigures
\message{If you do not have epsf.tex (to include figures),} \message{change the option at the top of the tex
file.}
\input epsf
\def\figin{\epsfcheck\figin}\def\figins{\epsfcheck\figins}
\def\epsfcheck{\ifx\epsfbox\UnDeFiNeD
\message{(NO epsf.tex, FIGURES WILL BE IGNORED)}
\gdef\figin##1{\vskip2in}\gdef\figins##1{\hskip.5in}
\else\message{(FIGURES WILL BE INCLUDED)}%
\gdef\figin##1{##1}\gdef\figins##1{##1}\fi}
\def\DefWarn#1{}
\def\figinsert{\goodbreak\midinsert}
\def\ifig#1#2#3{\DefWarn#1\xdef#1{Fig.~\the\figno}
\writedef{#1\leftbracket Fig.\noexpand~\the\figno}%
\figinsert\figin{\centerline{#3}}\medskip\centerline{\vbox{ \baselineskip12pt\advance\hsize by -1truein
\noindent\footnotefont{\bf Fig.~\the\figno:} #2}}
\bigskip\endinsert\global\advance\figno by1}
\else
\def\ifig#1#2#3{\xdef#1{Fig.~\the\figno}
\writedef{#1\leftbracket Fig.\noexpand~\the\figno}%
\global\advance\figno by1} \fi

\def\doublefig#1#2#3#4{\DefWarn#1\xdef#1{Fig.~\the\figno}
\writedef{#1\leftbracket Fig.\noexpand~\the\figno}%
\figinsert\figin{\centerline{#3\hskip1.0cm#4}}\medskip\centerline{\vbox{ \baselineskip12pt\advance\hsize by
-1truein \noindent\footnotefont{\bf Fig.~\the\figno:} #2}}
\bigskip\endinsert\global\advance\figno by1}

\useblackboard
\message{If you do not have msbm (blackboard bold) fonts,} \message{change the option at the top of the tex
file.} \font\blackboard=msbm10 scaled \magstep1 \font\blackboards=msbm7 \font\blackboardss=msbm5
\textfont\black=\blackboard \scriptfont\black=\blackboards \scriptscriptfont\black=\blackboardss

\else

\fi
%
\def\subsubsec#1{\bigskip\noindent{\it{#1}} \bigskip}
\def\yboxit#1#2{\vbox{\hrule height #1 \hbox{\vrule width #1
\vbox{#2}\vrule width #1 }\hrule height #1 }}
\def\fillbox#1{\hbox to #1{\vbox to #1{\vfil}\hfil}}
\def\ybox{{\lower 1.3pt \yboxit{0.4pt}{\fillbox{8pt}}\hskip-0.2pt}}
%
%


\def\comments#1{}



\def\II{\relax{I\kern-.10em I}}

\def\IZ{\relax{\rm Z\kern-.34em Z}}
\def\IB{\relax{\rm I\kern-.18em B}}
\def\IC{{\relax\hbox{$\inbar\kern-.3em{\rm C}$}}}
\def\ID{\relax{\rm I\kern-.18em D}}
\def\IE{\relax{\rm I\kern-.18em E}}
\def\IF{\relax{\rm I\kern-.18em F}}
\def\IG{\relax\hbox{$\inbar\kern-.3em{\rm G}$}}
\def\IGa{\relax\hbox{${\rm I}\kern-.18em\Gamma$}}
\def\IH{\relax{\rm I\kern-.18em H}}
\def\II{\relax{\rm I\kern-.18em I}}
\def\IK{\relax{\rm I\kern-.18em K}}
\def\IP{\relax{\rm I\kern-.18em P}}

%

\def\inbar{\,\vrule height1.5ex width.4pt depth0pt}

\def\IR{\relax{\rm I\kern-.18em R}}

\def\simgt{\hskip0.05in\relax{
\raise3.0pt\hbox{ $>$ {\lower5.0pt\hbox{\kern-1.05em $\sim$}} }} \hskip0.05in}

%


%

\def\lp10{\ell_p^{10}}
\def\lp11{\ell_p^{11}}
\def\R11{R_{11}}

\def\frac#1#2{{#1 \over #2}}



\newdimen\tableauside\tableauside=1.0ex
\newdimen\tableaurule\tableaurule=0.4pt
\newdimen\tableaustep
\def\phantomhrule#1{\hbox{\vbox to0pt{\hrule height\tableaurule width#1\vss}}}
\def\phantomvrule#1{\vbox{\hbox to0pt{\vrule width\tableaurule height#1\hss}}}
\def\sqr{\vbox{%
  \phantomhrule\tableaustep
  \hbox{\phantomvrule\tableaustep\kern\tableaustep\phantomvrule\tableaustep}%
  \hbox{\vbox{\phantomhrule\tableauside}\kern-\tableaurule}}}
\def\squares#1{\hbox{\count0=#1\noindent\loop\sqr
  \advance\count0 by-1 \ifnum\count0>0\repeat}}
\def\tableau#1{\vcenter{\offinterlineskip
  \tableaustep=\tableauside\advance\tableaustep by-\tableaurule
  \kern\normallineskip\hbox
    {\kern\normallineskip\vbox
      {\gettableau#1 0 }%
     \kern\normallineskip\kern\tableaurule}%
  \kern\normallineskip\kern\tableaurule}}
\def\gettableau#1 {\ifnum#1=0\let\next=\null\else
  \squares{#1}\let\next=\gettableau\fi\next}

\tableauside=1.0ex \tableaurule=0.4pt


 %
 %
 \def\eqnn#1{\xdef #1{(\secsym\the\meqno)}\writedef{#1\leftbracket#1}%
 \global\advance\meqno by1\wrlabeL#1}
 \def\eqna#1{\xdef #1##1{\hbox{$(\secsym\the\meqno##1)$}}
 \writedef{#1\numbersign1\leftbracket#1{\numbersign1}}%
 \global\advance\meqno by1\wrlabeL{#1$\{\}$}}
 \def\eqn#1#2{\xdef #1{(\secsym\the\meqno)}\writedef{#1\leftbracket#1}%
 \global\advance\meqno by1$$#2\eqno#1\eqlabeL#1$$}

\global\newcount\itemno \global\itemno=0

\def\itemaut#1{\global\advance\itemno by1\noindent\item{\the\itemno.}#1}


\def\eg{{\it e.g.}}

\hyphenation{Di-men-sion-al}


\lref\APS{
  A.~Adams, J.~Polchinski and E.~Silverstein,
  ``Don't panic! Closed string tachyons in ALE space-times,''
  JHEP {\bf 0110}, 029 (2001)
  [arXiv:hep-th/0108075].
}

\lref\scalings{
  S.~B.~Giddings,
  ``The fate of four dimensions,''
  Phys.\ Rev.\ D {\bf 68}, 026006 (2003)
  [arXiv:hep-th/0303031].
  E.~Silverstein,
  ``TASI / PiTP / ISS lectures on moduli and microphysics,''
  arXiv:hep-th/0405068.
}


\lref\SchmidhuberBV{
  C.~Schmidhuber and A.~A.~Tseytlin,
  ``On string cosmology and the RG flow in 2-d field theory,''
  Nucl.\ Phys.\ B {\bf 426}, 187 (1994)
  [arXiv:hep-th/9404180].
}

\lref\CooperVG{
  A.~R.~Cooper, L.~Susskind and L.~Thorlacius,
  ``Two-dimensional quantum cosmology,''
  Nucl.\ Phys.\ B {\bf 363}, 132 (1991)
}

\lref\HarveyWM{
  J.~A.~Harvey, D.~Kutasov, E.~J.~Martinec and G.~W.~Moore,
  ``Localized tachyons and RG flows,''
  arXiv:hep-th/0111154.
%
}

\lref\SCsemiclass{
J.~Polchinski, ``A Two-Dimensional Model For Quantum Gravity,'' Nucl.\ Phys.\ B {\bf 324}, 123 (1989);
B.~C.~Da Cunha and E.~J.~Martinec, ``Closed string tachyon condensation and worldsheet inflation,'' Phys.\ Rev.\
D {\bf 68}, 063502 (2003) [arXiv:hep-th/0303087];
E.~J.~Martinec, ``The annular report on non-critical string theory,'' arXiv:hep-th/0305148.
  S.~Hellerman and X.~Liu,
  ``Dynamical dimension change in supercritical string theory,''
  arXiv:hep-th/0409071.
 }

\lref\TE{
  J.~McGreevy and E.~Silverstein,
  ``The tachyon at the end of the universe,''
  JHEP {\bf 0508}, 090 (2005)
  [arXiv:hep-th/0506130].
}

\lref\juanooguri{
  J.~M.~Maldacena, H.~Ooguri and J.~Son,
  ``Strings in AdS(3) and the SL(2,R) WZW model. II: Euclidean black hole,''
  J.\ Math.\ Phys.\  {\bf 42}, 2961 (2001)
  [arXiv:hep-th/0005183].
}

\lref\seiberg{
  A.~Giveon, D.~Kutasov and N.~Seiberg,
  ``Comments on string theory on AdS(3),''
  Adv.\ Theor.\ Math.\ Phys.\  {\bf 2}, 733 (1998)
  [arXiv:hep-th/9806194].
}

\lref\het{
  D.~Orlando,
  ``AdS(2) x S**2 as an exact heterotic string background,''
  arXiv:hep-th/0502213.
  D.~Israel, C.~Kounnas, D.~Orlando and P.~M.~Petropoulos,
  ``Heterotic strings on homogeneous spaces,''
  Fortsch.\ Phys.\  {\bf 53}, 1030 (2005)
  [arXiv:hep-th/0412220].
}

\lref\dealwisetal{
  S.~P.~de Alwis, J.~Polchinski and R.~Schimmrigk,
  ``Heterotic Strings With Tree Level Cosmological Constant,''
  Phys.\ Lett.\ B {\bf 218}, 449 (1989).
}

\lref\ceff{see $\eg$
  D.~Kutasov and N.~Seiberg,
  ``Number Of Degrees Of Freedom, Density Of States And Tachyons In String
  Theory And Cft,''
  Nucl.\ Phys.\ B {\bf 358}, 600 (1991).
}

\lref\RWrefs{see $\eg$
  D.~Mitchell and N.~Turok,
  ``Statistical Properties Of Cosmic Strings,''
  Nucl.\ Phys.\ B {\bf 294}, 1138 (1987).
}

\lref\mysterious{
  A.~Iqbal, A.~Neitzke and C.~Vafa,
  ``A mysterious duality,''
  Adv.\ Theor.\ Math.\ Phys.\  {\bf 5}, 769 (2002)
  [arXiv:hep-th/0111068].
}

\lref\RSsaltman{
  A.~Saltman and E.~Silverstein,
  ``A new handle on de Sitter compactifications,''
JHEP {\bf 0601}, 139 (2006)
  [arXiv:hep-th/0411271].
}

\lref\SimeonXiao{
  S.~Hellerman and X.~Liu,
  ``Dynamical dimension change in supercritical string theory,''
  arXiv:hep-th/0409071.
}

\lref\TFA{
  A.~Adams, X.~Liu, J.~McGreevy, A.~Saltman and E.~Silverstein,
  ``Things fall apart: Topology change from winding tachyons,''
  JHEP {\bf 0510}, 033 (2005)
  [arXiv:hep-th/0502021].
}

\lref\SCsemiclass{
J.~Polchinski, ``A Two-Dimensional Model For Quantum Gravity,'' Nucl.\ Phys.\ B {\bf 324}, 123 (1989);
B.~C.~Da Cunha and E.~J.~Martinec, ``Closed string tachyon condensation and worldsheet inflation,'' Phys.\ Rev.\
D {\bf 68}, 063502 (2003) [arXiv:hep-th/0303087];
E.~J.~Martinec, ``The annular report on non-critical string theory,'' arXiv:hep-th/0305148.
}

\lref\TE{
  J.~McGreevy and E.~Silverstein,
  ``The tachyon at the end of the universe,''
  JHEP {\bf 0508}, 090 (2005)
  [arXiv:hep-th/0506130].
}

\lref\Gary{
  G.~T.~Horowitz,
  ``Tachyon condensation and black strings,''
  JHEP {\bf 0508}, 091 (2005)
  [arXiv:hep-th/0506166];
  S.~F.~Ross,
  ``Winding tachyons in asymptotically supersymmetric black strings,''
  arXiv:hep-th/0509066.
}

\lref\crapsetal{
  B.~Craps, S.~Sethi and E.~P.~Verlinde,
  ``A matrix big bang,''
  arXiv:hep-th/0506180.
}

\lref\edphases{
  E.~Witten,
  ``Phases of N = 2 theories in two dimensions,''
  Nucl.\ Phys.\ B {\bf 403}, 159 (1993)
  [arXiv:hep-th/9301042].
}

\lref\hubsch{see $\eg$ the explanation in
  T.~Hubsch,
  ``Calabi-Yau manifolds: A Bestiary for physicists,''
}

\lref\StromTak{
  A.~Strominger and T.~Takayanagi,
  ``Correlators in timelike bulk Liouville theory,''
  Adv.\ Theor.\ Math.\ Phys.\  {\bf 7}, 369 (2003)
  [arXiv:hep-th/0303221].
}

\lref\SKJD{
  J.~Distler and S.~Kachru,
  ``(0,2) Landau-Ginzburg theory,''
  Nucl.\ Phys.\ B {\bf 413}, 213 (1994)
  [arXiv:hep-th/9309110].
}

\lref\dSCFT{
  A.~Strominger,
  ``Inflation and the dS/CFT correspondence,''
  JHEP {\bf 0111}, 049 (2001)
  [arXiv:hep-th/0110087].
}

\lref\dSdS{
  M.~Alishahiha, A.~Karch and E.~Silverstein,
  ``Hologravity,''
  JHEP {\bf 0506}, 028 (2005)
  [arXiv:hep-th/0504056].
}

\lref\landscape{
  R.~Bousso and J.~Polchinski,
  ``Quantization of four-form fluxes and dynamical neutralization of the
  cosmological constant,''
  JHEP {\bf 0006}, 006 (2000)
  [arXiv:hep-th/0004134];
  S.~B.~Giddings, S.~Kachru and J.~Polchinski,
  ``Hierarchies from fluxes in string compactifications,''
  Phys.\ Rev.\ D {\bf 66}, 106006 (2002)
  [arXiv:hep-th/0105097];
  E.~Silverstein,
  ``(A)dS backgrounds from asymmetric orientifolds,''
  arXiv:hep-th/0106209;
  A.~Maloney, E.~Silverstein and A.~Strominger,
  ``De Sitter space in noncritical string theory,''
  arXiv:hep-th/0205316;
  S.~Kachru, R.~Kallosh, A.~Linde and S.~P.~Trivedi,
  Phys.\ Rev.\ D {\bf 68}, 046005 (2003)
  [arXiv:hep-th/0301240].
{\it et seq.}}

\lref\BV{
  N.~L.~Balazs and A.~Voros,
  ``Chaos On The Pseudosphere,''
  Phys.\ Rept.\  {\bf 143}, 109 (1986).
}

\lref\AndyD{
  A.~Strominger,
  ``The Inverse Dimensional Expansion In Quantum Gravity,''
  Phys.\ Rev.\ D {\bf 24}, 3082 (1981).
}

\Title{\vbox{\baselineskip12pt \hbox{SU-ITP-05/26}\hbox{SLAC-PUB-11510}}} {\vbox{ \centerline{Dimensional
Mutation and Spacelike Singularities}
%
%
%
%
%
}}
\bigskip
\bigskip
\centerline{Eva Silverstein}
\bigskip
\centerline{{\it SLAC and Department of Physics, Stanford University, Stanford, CA 94305-4060}}
\bigskip
\bigskip
\noindent

I argue that string theory compactified on a Riemann surface crosses over at small volume to a higher
dimensional background of supercritical string theory.  Several concrete measures of the count of degrees of
freedom of the theory yield the consistent result that at finite volume, the effective dimensionality is
increased by an amount of order $2h/V$ for a surface of genus $h$ and volume $V$ in string units.  This arises
in part from an exponentially growing density of states of winding modes supported by the fundamental group, and
passes an interesting test of modular invariance. Further evidence for a plethora of examples with the spacelike
singularity replaced by a higher dimensional phase arises from the fact that the sigma model on a Riemann
surface can be naturally completed by many gauged linear sigma models, whose RG flows approximate time evolution
in the full string backgrounds arising from this in the limit of large dimensionality. In recent examples of
spacelike singularity resolution by tachyon condensation, the singularity is ultimately replaced by a phase with
all modes becoming heavy and decoupling. In the present case, the opposite behavior ensues:  more light degrees
of freedom arise in the small radius regime. We comment on the emerging zoology of cosmological singularities
that results.

\bigskip
\Date{October 2005}


\newsec{Dimensionality and Singularities}

Many timelike singularities are resolved in a way that involves new {\it light} degrees of freedom appearing at
the singularity. In spacelike singularities studied recently \refs{\TE,\Gary}, ordinary spacetime ends where a
tachyon background becomes important. The tachyon at first constitutes a new light mode in the system, which
goes beyond the spectrum evident in general relativity.  However its condensation then replaces the would-be
short-distance singularity with a phase where degrees of freedom ultimately become {\it heavy}
\refs{\StromTak,\TE}.  In this note, we will find strong indications that there is a whole zoo of possible
behaviors at cosmological spacelike singularities, including examples in which the general relativistic
singularity is replaced by a phase with more light degrees of freedom, in the form of a larger effective number
of dimensions.\foot{see \crapsetal\ for an interesting null singularity where a similar behavior obtains.}

\subsec{Density of States}

Like spacetime itself, the notion of dimensionality is a derived concept.  In the context of perturbative string
theory, a more precise diagnostic for the effective dimensionality is the effective matter central charge
$c_{eff}$, which determines the high energy density of states of the system.  In simple situations, such as flat
space or linear dilaton backgrounds, this corresponds to the number of dimensions in which strings can
oscillate, but it applies equally well to small target spaces where the geometrical description breaks down. In
appropriate circumstances \ceff, the worldsheet modular group relates $c_{eff}$ to the Zamalodchikov $c$
function appropriate to the worldsheet matter sector.

For example, in noncritical limits of string theory, the worldsheet Weyl anomaly conditions are equivalent to
the equations of motion following from a spacetime effective action with a string-frame effective potential
\dealwisetal
\eqn\Dpot{ {\cal V}_{d}^{(s)}\sim {1\over g_{eff}^2}{{D-D_c}\over l_s^d} }
Here $D_c$ is the critical dimension, $g_s$ is the string coupling, and $g_{eff}^2=g_s^2/V$ for a
compactification from $D$ to $d$ dimensions on a space of volume $Vl_s^{D-d}$. That is, the Weyl anomaly has a
contribution proportional to the central charge $D-D_c$ of the matter sector, which must be balanced against
other contributions (such as time dependence in the dilaton). Correspondingly, the string can independently
oscillate in $D$ directions (modulo the ghost contributions which effectively remove two dimensions) and the
high energy density of states grows exponentially with the mass times $\sqrt{D}$.

Now consider critical string theory on a target space which includes a genus $h$ Riemann surface component
$\Sigma$ of volume $V_\Sigma l_s^2$. Dimensional reduction to eight dimensions on a constant curvature surface
yields a string frame effective potential
\eqn\RSpot{ {\cal V}_8^{(s)}\sim {1\over g_{eff}^2}{{2h-2}\over {V_\Sigma l_s^8}} }

Comparing \Dpot\RSpot, we see that the Weyl anomaly for $g_{eff}$ scales like $D-D_c$ in the supercritical case
and like $(2h-2)/V_\Sigma$ in the Riemann surface case.  This suggests an effective central charge
\eqn\RSWeyl{c_{eff}^{Weyl}\sim {{2h-2}\over V_\Sigma}}

At large volume $V_\Sigma \to\infty$, the theory reverts to ten dimensional flat space, for which the central
charge (or $\hat c$ in the superstring) of the Riemann surface direction is 2.  This corresponds to the number
of dimensions in which the string can oscillate. As $V_\Sigma$ decreases, the effective central charge appearing
in the Weyl anomaly increases.  As the volume $V_\Sigma l_s^2$ approaches the string scale, this effective
central charge approaches $2h-2$. More precisely, when $V_\Sigma$ is of order $2h-2$, the curvature ${\cal
R}\sim (2h-2)/(V_\Sigma l_s^2)$ reaches the string scale, and the formulas \RSpot\RSWeyl\ can be significantly
corrected.  In any case, $c_{eff}^{Weyl}$ increases as the volume of the surface decreases, suggesting a
connection between negatively curved target spaces at small radius and supercritical strings.

Now, this increase in the central charge for smaller size might at first sight appear surprising, as it does not
correspond to the naive number (2) of dimensions in which the string can oscillate.  However, this counting also
arises from two other interrelated calculations.  First, applying the Selberg trace formula to our problem
allows us to calculate the density of states of the lowest winding modes supported by the fundamental group of
the surface (see \BV\ for a review).  This yields
\eqn\densSelberg{\rho_{winding}\propto e^{A ml_s\sqrt{2h/V}}}
where $m$ is the mass of the state and $A$ a constant of order 1.  At large volume, where this calculation is
directly under control, these states are subdominant to the ordinary oscillator modes in two dimensions.

The only general conclusion we can make is that the effective dimensionality increases as the volume shrinks.
However it is interesting to contemplate the behavior for very small volume.  If we consider the large genus
limit $h\gg1$ and imagine extrapolating \densSelberg\ to $V\sim 1$, there is a picturesque way to see how an
effective dimensionality of order $2h$ might emerge. This arises from recognizing the states supported by the
fundamental group as mapping out a lattice random walk in $2h$ dimensions, which could reach its continuum limit
if the corresponding modes persist to very small volume.

Let us label the cycles of the Riemann surface in a symplectic basis by $a_i, b_i$ with $i=1, \dots, h$, and
their inverses by $a_i^{-1}$, $b_i^{-1}$ (keeping track of orientation). The fundamental group of a Riemann
surface is given by ``words" formed from paths going around these cycles, subject to the inverse relations
$a_ia_i^{-1}=1=a_i^{-1}a_i$ and $b_i b_i^{-1}=1=b_i^{-1}b_i$ and the relation
\eqn\finalreln{\prod_{i=1}^h a_i b_i a_i^{-1} b_i^{-1} = 1}
Topologically stable string states arise for each independent element of this group. For appropriately shaped
Riemann surfaces, at large $h$, they correspond to a lattice random walk in a $2h$-dimensional space:  at each
step, the string can wind in either direction around any of the $2h-1$ available cycles (with the $-1$ here
arising from the inverse relations precluding a step taken in the opposite direction from the previous step).
Also at large $h$, the further relation \finalreln\ is a subleading effect.   This identification of the winding
states with an isotropic $2h$-dimensional random walk requires some conditions on the shape of the surface; for
example it requires all the cycles of the same order. The selection rules determined by the fundamental group
ensure that the states we are counting are independent states before string interactions are taken into account
which allow them to decay.  At large volume, these lattice random walk states are far from the continuum limit,
but at very small volume if these states persist they behave like strings in $2h$ dimensions.  In particular, at
high energies a typical string state in $D-D_c\sim 2h-2 \gg 1$ dimensions also behaves like a random walk in
$D\sim 2h$ dimensions \RWrefs.

In this correspondence, dimensionality mutates into topology\foot{a feature suggested previously in \mysterious\
in a different context}. In the next sections, we will elaborate on the physical context, interpretation, and
perturbative self-consistency of this proposal. In the absence of additional ingredients, the Riemann surface at
large radius evolves with time according to the vacuum Einstein equations.  We will focus on this case in \S3,
where the supercriticality provides an intriguing approach to resolving the associated cosmological singularity.

A more general setting in which to apply this connection is in a Riemann surface compactification with extra
ingredients helping to stabilize it, as in \RSsaltman.  This may slow or stop the time evolution in appropriate
cases and permit comparison of metastable backgrounds of the Riemann surface and of the supercritical theory
(itself with extra ingredients, perhaps along the lines of \landscape).  Other potential connections between
critical and supercritical backgrounds of string theory were suggested in \SimeonXiao, and between critical and
subcritical backgrounds in many works on closed string tachyon condensation.

\subsec{Zoology}

It is worth emphasizing that the value $2h$ for $\hat c_{eff}$ is not always attained as a Riemann surface
shrinks; this depends on its moduli and spin structure. For example, as a Riemann surface shrinks, it can
undergo other transitions \refs{\TFA,\TE,\Gary}\ which reduce its topology and/or separate it into distinct
components if the spin structure is chosen appropriately and/or the surface approaches a factorization limit.
Moreover, the cycles supporting the fundamental group elements might approach the string scale at different
rates and at different times, yielding a smaller high energy density of states arising from the corresponding
subgroup of the fundamental group.   In addition, {\it larger} values of the UV central charge are also possible
in sectors transverse to the Riemann surface; I will focus on the degrees of freedom that are identifiable on
the Riemann surface side of the transition.

A simple construction we will study in \S2\ and the Appendix reveals a large zoo of cosmological backgrounds
with the common property that dimensionality decreases and topology becomes more copious as time evolves.  We
will employ the relation of RG flow and time evolution in systems with extra supercritical dimensions
\refs{\SchmidhuberBV,\CooperVG}\ to obtain large classes of examples of FRW-like cosmologies with compact
Riemann surface spatial slices whose spacelike singularity is replaced by a phase of higher dimensional string
theory.  Not surprisingly, there is a large multiplicity of string cosmologies reminiscent of the large
multiplicity of metastable vacua \landscape.  Although this renders any one of them less compelling on its own,
it is worth making the qualitative point that spacelike singularities can be corrected by stringy effects in
both directions (more light degrees of freedom as in the present examples, versus fewer in the tachyon
condensation examples \TE\Gary).  In any case, only with a sufficiently broad understanding of the range of
early time behaviors can one hope to accurately formulate initial conditions in string theory.

\subsec{A Modular Invariance test}

In this subsection, I will use Euclidean AdS conformal sigma models and their compact orbifolds to perform a
concrete test of this proposal of an interpolation between negatively curved spaces with large fundamental group
and higher dimensional string target spaces.  As discussed above, in the spaces of interest, the density of
winding modes grows like $e^{ml_s\sqrt{2h/V}}$ \densSelberg.  These states are present only by virtue of the
compactness of the surface. By modular invariance, there must exist a corresponding light mode propagating in
the dual channel of the 1-loop partition function.

Consider the case that the compact Riemann surface (or higher dimensional analogue) is obtained by orbifolding a
Euclidean AdS space by a Fuchsian group $\Gamma\subset SL(2,R)$ (or higher dimensional analogue). Euclidean
$AdS_3$ worldsheet CFTs were studied in \refs{\seiberg,\juanooguri}\ and similar $AdS_2$ models in
compactifications of the heterotic string were studied in \het.  These Euclidean models involve imaginary fluxes
(as described for example in \seiberg) and no time dimension; hence their physical application is limited at
best. However, as we will see these models provide a formal setting in which it is simple to test whether the
possibility raised above could occur in this class of examples consistent with modular invariance.\foot{It
should be borne in mind however that the physical interpretation of these models after orbifolding is not clear;
before orbifolding they correspond to the gravity duals of Euclidean CFTs, but after orbifolding the boundary is
absent and this interpretation is lacking.  In any case we will find it useful as a formal exercise in
worldsheet conformal field theory.}

Consider a Riemann surface $\Sigma$ obtained by orbifolding Euclidean $AdS_2$.  The fundamental group is
nontrivial after orbifolding, and the winding strings are twisted states of this orbifold. As such, these states
obey the selection rules of the orbifold model and do not mix linearly with ordinary untwisted oscillator modes.
For an arbitrary Riemann surface, we do not know their mass spectrum at very small volume, and would like to
know if it is consistent to imagine that they might come to dominate the high energy density of states in this
regime in some subset of models.

There is a (too-)quick argument suggesting that this possibility is ruled out in orbifold models of the sort we
have set up.  One-loop modular invariance relates the high energy density of states to the dimension of the
lowest energy state propagating in the dual channel.\foot{See \HarveyWM\ for a recent review of this relation
and application to simpler models \APS\ with tachyons.} If the high energy density of states is to be enhanced
by the presence of the winding states, whose effective central charge grows like $h/V_\Sigma$ as the volume
decreases, then there has to be an effectively tachyonic mode in the spectrum whose effective mass squared
becomes more and more negative. More specifically, the relation $c_{eff}\sim 2h/V$ requires that the density of
states as a function of mass $m$ in the putative supercritical limit is of the form
\eqn\densSC{\rho(M)\sim e^{A\sqrt{2h/V}ml_s}\sim e^{A m l_s^2/R}}
where $A$ is an order 1 constant and $R$ the curvature radius of the surface.  By a modular transformation, this
translates to an effective mass squared for the lightest mode of order
\eqn\massneeded{m_T^2\sim -2h/V_\Sigma l_s^2\sim -1/R^2}

This mode does not exist in the parent theory, corresponding to the fact that in that theory the effective
central charge is of order 1.  The puzzle is to understand if the orbifolding could produce such a state. Short
winding strings can appear in twisted sectors of the orbifold in the standard way, and these can become
tachyonic if the fermion boundary conditions are chosen appropriately, but at least in the geometrical regime
the $R$-dependence of the winding string masses is different from the $-1/R^2$ behavior required.

\subsubsec{Resolution: the Volume Mode}

Euclidean $AdS_3$ supported by (imaginary \seiberg) Neveu-Schwarz 3-form flux may be orbifolded to obtain a
compact hyperbolic 3-manifold with nontrivial fundamental group.  This would share some of the features of the
examples of interest, and leads to the puzzle just noted.  The same puzzle arises in the simpler case of
Euclidean $AdS_2$, supported by (imaginary) 2-form flux, which can be obtained for example in the heterotic
theory \het. This can be orbifolded to obtain a compact Riemann surface.  The resolution of the puzzle raised
above in both these cases is simply that the volume modulus of the compact surface obtained by orbifolding is
tachyonic, with the right scaling to match the requirement \massneeded.


Let us begin with a parent theory with no tachyonic IR divergences in its partition function. See for example
the tour de force calculations in \juanooguri\ for explicit formulae in the analogous Euclidean $AdS_3$
examples. All the tachyons in the parent theory are of the ``allowed" variety according to the
Breitenlohner-Freedman bound, stabilized by the boundary conditions of the noncompact Euclidean AdS.

Next let us consider the orbifolded theory, a compact constant curvature Riemann surface $\Sigma$ with heterotic
2-form flux $F\sim iQ_1/V_\Sigma$. It is compact, with no boundary. Hence in addition to the twisted winding
strings noted above, this theory also has a dynamical volume modulus $V_\Sigma \equiv e^{2\gamma_1}$ (in our
convention $V_\Sigma$ is dimensionless, the physical volume being $V_\Sigma l_s^2$).


In fact there is a tachyon in the spectrum, which arises for a simple reason.  Dimensionally reducing along
$\Sigma$, the string-frame Euclidean effective action for the volume mode is of the form (up to order 1
coefficients)
\eqn\action{{\cal S}_\gamma = \int {{d^8 x}\over {l_s^8 g_s^2} }\sqrt{G_8} e^{2\gamma_1}\biggl(
|\nabla\gamma_1|^2 + \bigl[(2h-2) e^{-2\gamma_1} -{Q_1^2 e^{-4\gamma_1}}\bigr] \biggr) }
(See \scalings\ for similar analyses in the case of real-flux vacua.) In particular, $\gamma_1$ has an effective
potential (proportional to the term in square brackets) with a positive term from the negative curvature of
$\Sigma$, and a negative term from the the kinetic term for the imaginary flux. In the $\gamma_1$ direction, the
system is perched atop an unstable potential.
The orbifold CFT corresponds to the unstable maximum of this potential, at $e^{2\gamma_{1,max}}\sim
Q_1^2/(h-1)$. The fluctuation of $\gamma$ about this maximum is a tachyon.  Its mass squared is of order
\eqn\massresult{m_\gamma^2 \sim -{1\over R^2} \sim -{(2h-2)\over{ V_\Sigma l_s^2}} }
This provides a mode with the scaling required \massneeded\ for the test of modular invariance. In particular,
the partition function will have a corresponding IR divergence, in contrast to the parent Euclidean AdS theory
whose tachyons are all allowed and do not yield divergences \juanooguri.

Note also that similar considerations may apply to other moduli generated by the orbifold procedure.  We focus
here on the volume mode because it is universally present in compact examples.  This correlates with the fact
that the winding strings discussed above \densSC\ are present in the compact examples but not in the parent
Euclidean AdS theory.

\newsec{Renormalization Group Interlude}

In order to formulate perturbative string theory on a negatively curved target space, we must specify a well
defined sigma model.  The nonlinear sigma model on a negatively curved target space is strongly coupled in the
ultraviolet, and requires a UV completion.\foot{At the level of effective field theory of course one may
alternatively define the sigma model with a cutoff.}   Such a UV completion is easily obtained at the level of
the two-dimensional matter sigma model by embedding the Riemann surface into a higher dimensional space swept
out by some scalar fields $Y^i$. A potential ${\cal U}(Y)$ restricting the $Y$ fields to the Riemann surface is
a relevant perturbation of the UV fixed point.  In the UV, its effects go away, and the system lives on the
space swept out by the $Y$ fields (or more generally an abstract CFT with a central charge $\hat c_{UV}>2$).  In
the IR, the potential produces a nonlinear sigma model on the Riemann surface.

Of course in a generic theory, the potential ${\cal U}(Y)$ may be strongly corrected upon flow to the IR, so
that fixing to the desired potential energy and the corresponding RG flow to the Riemann surface nonlinear sigma
model would require significant fine tuning.  However, in the case of worldsheet $(2,2)$ or $(0,2)$
supersymmetry, such corrections can be straightforwardly controlled. In particular, using the standard
technology of gauged linear sigma models \edphases, it is possible to identify supercritical CFTs which provide
the required UV fixed points in many examples, as I review in the appendix.  These examples include cases where
a supercritical analogue of the type II GSO projection is available to the system, limiting the instabilities to
which the system is susceptible as compared to more general analogues of the type 0 GSO projection.

As discussed in \S1, the fundamental group of a Riemann surface of genus $h$ can support an effective central
charge as high as $2h$ for $V\to 1$.  However, the embedding spaces involved in the linear sigma model procedure
for formulating the sigma model need not be of dimension $2h$ (and need not be geometrical at the UV fixed
point) though they all have UV central charge greater than $\hat c=2$. This corresponds to the fact discussed
above that in order to realize the maximal $2h$ dimensional effective central charge supportable by the
fundamental group, the surface must be rather symmetrical, and tachyonic modes need to be avoided, as the
surface approaches the string scale. More generally, there are many consistent early time behaviors that arise
from this class of constructions.

This linear sigma model definition of the sigma model on a Riemann surface does not in itself guarantee that
time dependent backgrounds exist realizing an evolution between the supercritical and Riemann surface target
spaces. However at large effective central charge the worlsheet matter RG is a good approximation to the
spacetime evolution, as we will review and apply in the next section.

\newsec{Negative Curvature and Spacelike Singularities}

\subsec{RG flow and time evolution at large $c_{eff}$}

A higher genus Riemann surface, being negatively curved, is not by itself a static solution to string theory. At
large radius, in the absence of extra stress energy sources its scale factor $a(t)$ evolves according to the
vacuum Friedmann equation
\eqn\friedmann{\left({\dot a\over a}\right)^2= {1\over a^2} }
which forces the surface to expand or contract as a function of time according to
\eqn\GRevolution{a(t)=\pm (t-t_0)}
When $t$ reaches $t_0$, there is a spacelike singularity.  Here for simplicity we consider constant spatial
curvature; with more general choices the curvature perturbations smooth out at large $|t-t_0|$ but may be
significant at small $|t-t_0|$.

A nonlinear sigma model on a Riemann surface target space flows toward large radius in the infrared.  In string
theory, the solution \friedmann\GRevolution\ provides the correct dressing of this flow by worldsheet gravity to
obtain a consistent background in the large radius regime.  As the Riemann surface shrinks as $t$ tends toward
$t_0$ (either in the past or the future depending on the sign \GRevolution), the effective central charge of the
matter sigma model increases as discussed in the previous sections. This contrasts to the case of positively
curved spatial slices, and the similar case of flat spatial slices with winding tachyon modes, where the
singularity is associated with a lifting of degrees of freedom, a decrease in $c_{eff}$ \TFA\TE\Gary.\foot{A
null singularity where the flow of degrees of freedom is as in the present case was suggested in \crapsetal.}

It is difficult in general to follow the time evolution of this system through the string scale.  However, the
renormalization group flow in the matter sigma model becomes a good approximation to the time evolution in the
case that the effective central charge is large \refs{\SchmidhuberBV,\CooperVG}. One way to see this is as
follows. Consider an RG flow from a UV fixed point to an expanding Riemann surface, such as that determined by
the gauged linear sigma model in the previous section. The worldsheet action is semiclassically
\eqn\wsac{{\cal S}_{ws}={\cal S}_{UV} + \int d^2\sigma \rho {\cal O} e^{\kappa X^0}\pm \int {\cal
R}^{(2)}c_{eff}X^0}
where ${\cal O}$ is the relevant operator by which the UV CFT is perturbed at the beginning of the flow, and
$\kappa$ is determined by the requirement that the operator ${\cal O}e^{\kappa X^0}$ be marginal. The scale
transformations have a classical contribution arising from the worldsheet curvature term, and in the limit of
large $c_{eff}$, the two solutions for $\kappa$ scale like $\mp 1/\sqrt{c_{eff}}$ and $\pm\sqrt{c_{eff}}$.  To
fix the signs, consider a dilaton evolving from strong to weak coupling.  Then for fixed ${\cal O}$ dimension
$\Delta$, the growing mode of the tachyon has $\kappa\propto 1/\sqrt{c_{eff}}$.

In the cases with $\kappa\propto\pm1/\sqrt{c_{eff}}$, the contributions to the worldsheet path integral coming
from fluctuations of $X^0$ itself are suppressed in this limit \refs{\SCsemiclass,\SimeonXiao}. The matter
fluctuations contribute without any such suppression, and generate the RG flow of the theory, with the IR
behavior corresponding to long times (large values of the $X^0$ zero mode $X^0_0$).

For this reason we will mostly focus on the case where the effective central charge is large as the system
evolves from its supercritical description to a Riemann surface with string scale cycles, $V_\Sigma \sim (2h-2)
l_s^2$. As the surface further evolves to larger radius, the central charge of the Riemann surface component
drops closer to 2, and the confluence of worldsheet RG and time evolution does not remain true in the case of
the Riemann surface compactication of critical string theory, though the endpoints may still agree. However, the
evolution through the transition is under control in the case where we include a large spectator supercritical
sector to slow down the evolution.

\subsec{Coupling}

Given a crossover to a supercritical phase, the large $c_{eff}$ of the system sources the dilaton. An important
question is the behavior of the dilaton during the crossover, and the direction of evolution of the string
coupling in the supercritical phase.

The transition timescale is finite, of order $1/\kappa$.  The dilaton can be shifted
\eqn\dilshift{\Phi\to\Phi-\Phi_0}
by a large constant $\Phi_0$, to obtain an arbitrarily weak coupling during the transitional epoch between the
higher dimensional and Riemann surface phases.

However, to obtain a controlled description, one must address the behavior of the dilaton in the far past
(fixing to the big bang case where the Riemann surface expands in the far future) while addressing as in \S3.1\
the rate of time evolution in the background.  Let us consider two cases in turn:

\noindent 1)  If we choose the signs suggested in \S3.1, in which the tachyon describing the flow from the
supercritical theory to the Riemann surface theory turns on slowly according to $\kappa\propto
1/\sqrt{c_{eff}}$, then the string coupling evolves toward weak coupling as time evolves forward.  This means
that although the transition itself occurs during an epoch of weak coupling using \dilshift, in the far past the
system would be strongly coupled in the absence of other ingredients.  In the presence of stabilizing
ingredients as in \landscape, we may shut off this strong coupling behavior by metastabilizing the dilaton.
Alternatively, one may consider including transverse degrees of freedom and a second tachyon field which grows
toward the past and turns off by the time of the transition of interest; this however must turn off rapidly,
which may lead to strong back reaction.

\noindent 2)  Conversely, we may choose the signs so that the dilaton is weakly coupled in the past, but this
introduces a rapid time dependence in a generic tachyon which effects the transition.  One could tune to obtain
a very nearly marginal operator ${\cal O}$ in \wsac\ to slow it down.

\newsec{Discussion}

The phenomena discussed in this work suggest a role for supercritical string theory near a spacelike singularity
for compact negatively curved FRW , studied here in the case of compactification down to a 2+1 dimensional FRW
solution with Riemann surfaces as the spatial slices.  Although we focused here on the simplest negatively
curved spaces, Riemann surfaces, the methods would generalize to hyperbolic 3-manifolds with large fundamental
group.

These systems exhibit time dependent transitions from supercritical dimensionality to copious topology and lower
dimensionality. The evolution between higher and lower effective dimensionality is a promising trend, as it
provides a possibility of a dynamical determination of the dimensionality.

In the spacelike singularities studied in \TE, and likely in similar cases with positively curved spatial
slices, the singularity is replaced by a phase of tachyon condensate lifting closed string degrees of freedom,
leading in the big bang case to an evolution as time moves forward from fewer to more degrees of freedom as
measured by $c_{eff}$. Similarly, as discussed in \dSCFT\dSdS, the evolution in realistic inflationary cosmology
tends in this direction in terms of the evolution of the de Sitter entropy. The present set of examples works in
the opposite way, with more degrees of freedom (as measured by $c_{eff}$) appearing in the phase replacing the
singularity. Both starting points are interesting to consider for the question of the space of initial (or
final) conditions and its measure in light of the string landscape of late time solutions \landscape.

Another intriguing aspect of the present study is the role of the simplifications that appear at large
$c_{eff}$: this aided our analysis of the high energy density of states and provided a relation between the RG
and time dependent evolution. In other corners of M theory the simplifications arising in the presence of large
numbers of degrees of freedom have played an important role.  The limit of large dimensionality might play a
simplifying role in a way that is particularly suited to time dependent backgrounds and string-theoretic
cosmology.\foot{Simplifications at large $D$ have also appeared in \landscape\AndyD\SimeonXiao\SCsemiclass.}

Finally, the study of string target spaces at small radius has a rich tradition of connecting to mathematics.
The case of broken supersymmetry is likely to continue this confluence, with techniques such as surgery \TFA\
realized naturally in string theory, and generalizations of the CY-LG correspondence playing a role as discussed
here.  It would also be interesting if the present considerations could help clarify the physical context
appropriate for interpreting \mysterious.

\bigskip

\noindent{\bf Acknowledgements}

I would like to thank A. Adams, S. Kachru, G. Horowitz, X. Liu, A. Maloney, J. McGreevy, J. Polchinski, S.
Shenker, and A. Tomasiello for very useful discussions.  I thank the KITP for hospitality during part of this
work.  This research is supported in part by the DOE under contract DE-AC03-76SF00515 and by the NSF under
contract 9870115.

\appendix{A}{}

In this appendix, I explore two gauged linear sigma models realizing supercritical UV fixed points which flow
upon relevant perturbation to nonlinear sigma models on Riemann surfaces.  These models illustrate the process
discussed in \S3:  in the presence of a large supercritical spectator sector, the RG flow of the worldsheet
matter sector approaches time evolution in the spacetime background, and so the flow from an embedding space to
the nonlinear sigma model provides a time dependent background.  Since Riemann surface sigma models require a UV
completion with more degrees of freedom, the fact that the small volume (spacelike singular) region is replaced
by a phase with more degrees of freedom is generic.  Of course there are many different ways this can happen in
detail, as the following examples will illustrate.

Perhaps the simplest linear sigma model flowing to a Riemann surface nonlinear sigma model is obtained by
realizing the surface via a degree $n$ algebraic defining equation in $\IC\IP^2$.  In the notation and
conventions of \edphases, this is obtained in a $(2,2)$ gauged linear sigma model with the following field
content and interactions.  Three chiral multiplets $\Phi_i$ of charge $1$ under a single $U(1)$ gauge symmetry
alone would yield the $(2,2)$ $\IC\IP^2$ model.  If we add another chiral multiplet $P$ with charge $-n$, the
potential energy of the model is determined by the $D$-term and superpotential terms
\eqn\Dterm{D^2=(-n|P|^2+\sum_{i=1}^3 |\Phi_i|^2-\rho)^2}
\eqn\Fterm{W=P G_n(\Phi)}
where $G_n$ is a degree $n$ polynomial in the $\Phi$ fields and $\rho$ is the Fayet-Iliopoulos term of the
model. The infrared physics is dominated by the space of field configurations minimizing the potential energy.
At large positive $\rho$, the D-term and gauge invariance restrict the $\Phi$'s to lie on $\IC\IP^2$ (given a
sufficiently generic polynomial $G_n$ which requires $P=0$ for any nonzero $\Phi_i$).  The superpotential
requires $G_n(\Phi)= 0$, further restricting the fields to trace out a one complex dimensional locus.\foot{For
most values of $n$, there are also extra massive vacua as described in \edphases.} This surface is a two-torus
$T^2$ if $n=3$.

The running of the coupling $\rho$ is determined in terms of the sum of the charges as
\eqn\runningrho{{d\rho\over{d ln(\mu)}}=-n+3}
For $n=3$, $\rho$ is marginal.  For $n > 3$, $\rho$ runs to $+\infty$ in the IR and $-\infty$ in the UV, flowing
toward large volume in the IR as expected for a negative curvature target space. The genus $h$ of the surface in
the large radius geometrical phase is
\eqn\genusI{2-2h = n(3-n)}
In this geometrical phase applicable in the deep IR, the effective central charge approaches $\hat c=2$ (where
our ``hat" notation is defined such that $\hat c=1$ corresponds to a single real dimension).

In the UV, $\rho\to -\infty$, and the model reduces to a Landau-Ginzburg orbifold.  This model has central
charge
\eqn\cUVI{\hat c_{UV}=6\times (1-{2\over n})>2}

Altogether, this model describes a flow starting from a UV CFT given by a Landau-Ginzburg orbifold. Perturbing
by the relevant operator corresonding to $\rho$ yields a flow to a nonlinear sigma model on a Riemann surface.
This provides a UV complete sigma model for the Riemann surface target space, which exhibits explicitly a
decrease of the $c$ function.

As discussed in \TFA, this model is consistent with a chiral GSO projection if the sum of the charges $3-n$ is
even, providing a non-anomalous $\IZ_2$ subgroup of the chiral R symmetry.

This model, while supercritical in the UV, does not realize the full $2h$ dimensional target space that would be
obtained in cases where the winding states discussed in \S1\ persist.\foot{As discussed in  \S1, decay processes
\TFA\ can occur at the string scale, depending on the moduli and spin structure.}

We will next consider a slightly more complicated model which realizes a flow from a UV fixed point with a much
higher UV central charge, $c\propto h$ for large $h$, to a Riemann surface of genus $h$.  Again in the notation
and conventions of \edphases, consider the following $(2,2)$ symmetric gauged linear sigma model describing a
complete intersection of hypersurfaces embedded in $(\IC\IP^2)^m$.  Start with a gauge group $U(1)^m$, and
include for each factor in the gauge group 3 chiral multiplets of charge 1 (and uncharged under the remaining
factors).  That is, consider chiral multiplets $\Phi_i^{r}$ with $r=1,\dots, m$ indexing the $U(1)$ factor under
which $\Phi$ is charged, and with $i=1,2,3$ labeling the 3 fields of this charge.  Add negatively charged chiral
fields $P_r$ and $P_{r,r+1}$ as follows:
\eqn\Pcharges{\eqalign{& P_r  ~~~ r=1,\dots, m ~~~~ {\rm charge ~~ -3 ~~ under} ~~ U(1)_r \cr & P_{r, r+1} ~~~
r=1,\dots, m-1 ~~~ {\rm charge ~~ -1 ~~ under} ~~ U(1)_r ~~ {\rm and} ~~ U(1)_{r+1}\cr }}

The corresponding $D$ terms of the model are of the form
\eqn\Dterms{D_r^2=(-3|P_r|^2-|P_{r, r+1}|^2-|P_{r-1,r}|^2 +\sum_{i=1}^3|\Phi_i|^2-\rho_r )^2}
for $r=2,\dots, m-1$ (while for $r=1$ and $m$ one of the $|P_{r, r+1}|^2$ terms is absent). The sum of the
charges being negative translates into the statement that the Fayet-Iliopoulos couplings $\rho_r$ run in the
same direction as in our previous model:  they run to large positive values $\rho_r\to +\infty$ in the IR, and
$\rho_r\to -\infty$ in the UV.

At large positive $\rho$, the D-terms and gauge invariance restrict the $\Phi$ fields to lie on $(\IC\IP^2)^m$.
As in \Fterm\ above, superpotential terms and hence algebraic defining equations for our surface arise from
gauge invariant terms of the form $P G(\Phi)$, with the charges of the $P$ fields \Pcharges\ determining the
degrees of the defining equations.  The first set of $m$ equations corresponding to the $P_r$ \Pcharges\ alone
further restrict the $\Phi$ fields to live on $(T^2)^m$. Including the remaining set of $m-1$ equations yields
enough independent equations to restrict the model to a one complex dimensional space. Because $\rho_r$ grow
toward the IR, this surface grows in size in the IR and hence is a higher genus Riemann surface.

The adjunction formula \hubsch\ for the first Chern class again provides a simple way to compute the genus of
this surface: it is given by
\eqn\genusII{\int c_1 = 2-2h = 6-6m \Rightarrow h = 3m-2 }
which reduces at large genus to $2h\sim 6m$.

In this model the UV theory arises for $\rho\to -\infty$.  In the first model discussed above, this UV CFT was a
Landau-Ginzburg model.  In the present model, is a hybrid phase in the language of \edphases.  It has a UV
central charge proportional to $h$, as can be seen as follows.  Some $P$ fields in each $D$ term \Dterms\ must
be turned on.  The zeroes of the potential then occur for $\Phi^r_i=0$ but for massless $\Phi$ fields their
fluctuations about the minima yield contributions to the central charge as in Landau-Ginzburg theory.  Consider
for example the regime where $P_{r, r+1}\ne 0$.  The corresponding superpotential terms mass up two of the three
linear combinations of $\Phi^r$ fields for each $r$.  If $P_r$ are also nonzero, the remaining $\Phi$ fields
live on a Landau-Ginzburg theory with central charge $2m/3$.  This is an underestimate for the UV central
charge, as the $P$ fields can independently fluctuate to some extent.  This hybrid phase is somewhat
complicated; in analogous heterotic models a simpler phase arises in similar circumstances \SKJD. The hybrid
phase is somewhat complicated, so it is worth noting that in heterotic versions of this model, an LG phase
rather than a hybrid phase can be obtained as in \SKJD.  This occurs because in $(0,2)$ supersymmetry the
defining equations arising in the $(0,2)$ superpotential terms need not involve extra scalar fields $P$ but
instead multiply left-moving fermion superfields.   In this model, at large $2h\sim 6m$ the UV central charge is
proportional to $h$.

Of course there are many variants of this, including ones in which other spectator sectors appear.  It would be
interesting to map the winding modes more closely to the linear sigma model variables, tuning couplings to
preserve the winding modes in the $V\to 1$ limit of interest in \S1.  The description of the fundamental group
and winding modes in gauged linear sigma models is generally more complicated than describing the flow, though
some steps in this direction were taken in \TFA.

\listrefs

\end